\documentclass[letterpaper, 10pt, conference]{ieeeconf}
\usepackage[utf8]{inputenc}
\usepackage{amsmath}
\usepackage{amssymb}
\usepackage{hyperref}

\usepackage{amsthm}
\usepackage{xcolor}
\usepackage{graphicx}
\usepackage{svg}
\usepackage{arydshln}
\usepackage[ruled,vlined]{algorithm2e}
\usepackage{algpseudocode}
\usepackage{multirow}

\newtheorem*{remark}{Remark}

\newcounter{daggerfootnote}
\newcommand*{\daggerfootnote}[1]{%
    \setcounter{daggerfootnote}{\value{footnote}}%
    \renewcommand*{\thefootnote}{\fnsymbol{footnote}}%
    \footnote[2]{#1}%
    \setcounter{footnote}{\value{daggerfootnote}}%
    \renewcommand*{\thefootnote}{\arabic{footnote}}%
    }

\IEEEoverridecommandlockouts
\title{\LARGE \bf{Decentralized Role Assignment in Multi-Agent Teams via Empirical Game-Theoretic Analysis}}
\author{Fengjun Yang$^{1}$, Negar Mehr$^{2}$, and Mac Schwager$^{3}$
\thanks{Toyota Research Institute (``TRI")  provided funds to assist the authors with their research but this article solely reflects the opinions and conclusions of its authors and not TRI or any other Toyota entity. This work was also supported in part by ONR grant number N00014-18-1-2830.}
\thanks{$^{1}$Fengjun Yang is with the Department of Computer and Information Science, the University of Pennsylvania, Philadelphia, PA 19104 USA
        {\tt\small fengjun@seas.upenn.edu}}
\thanks{$^{2}$Negar Mehr is with the Aerospace Engineering Department, the University of Illinois at Urbana-Champaign, Urbana, IL 61801 USA
        {\tt\small negar@illinois.edu}}
\thanks{$^{3}$Mac Schwager is with the Department of Aeronautics and Astronautics, Stanford University, Stanford, CA 94305 USA
        {\tt\small  schwager@stanford.edu}}%
}

\begin{document}

\maketitle
\thispagestyle{empty}
\pagestyle{empty}

\begin{abstract}
We propose a method, based on empirical game theory, for a robot operating as part of a team to choose its role within the team without explicitly communicating with team members, by leveraging its knowledge about the team structure. To do this, we formulate the role assignment problem as a dynamic game, and borrow tools from empirical game-theoretic analysis to analyze such games. Based on this game-theoretic formulation, we propose a distributed controller for each robot to dynamically decide on the best role to take. We demonstrate our method in simulations of a collaborative planar manipulation scenario in which each agent chooses from a set of feedback control policies at each instant.  The agents can effectively collaborate without communication to manipulate the object while also avoiding collisions using our method.

\end{abstract}

\section{Introduction}
Roles arise naturally as a way to divide responsibilities and facilitate collaboration on human teams. By thinking in terms of roles, a team of humans can efficiently make plans to coordinate themselves without reasoning explicitly about specific low-level tasks. Role structures, however, are not unique to human teams. For example, on a robot soccer team, one would similarly consider dividing the robots into goalies, defenders, attackers, etc. \cite{playne2008knowledge}. In this work, we study how we can coordinate a heterogeneous team of robots to collaborate on time-extended tasks by choosing an ideal sequence of roles for the robots to take. Further, to achieve better scalability and robustness against single-point failures, we build a decentralized algorithm where each robot autonomously decides on its own sequence of roles.

Our problem of role assignment is closely related to that of task allocation
\daggerfootnote{
While both roles and tasks encode responsibilities and actions (some may indeed argue that they are equivalent \cite{gerkey2003role}), we here choose to call our problem role assignment for two reasons. First, we consider time-extended tasks. The choice of ``role" stresses that the responsibilities allocated to a robot are in effect for an extended period of time. Secondly, it signals that a robot can adaptively switch roles online as it see fits; this is to be contrasted with tasks, which usually need to be completed before a new one can start.
}
, where one has to allocate a set of tasks to a team of robots in a way that maximizes the overall performance of the team \cite{gerkey2004formal, korsah2013comprehensive, khamis2015multi}. Within the task allocation literature, our work is most closely related to market-based task allocation algorithms \cite{gerkey2002sold, dias2004traderbots, zlot2006auction}. In these algorithms, a tasks is broadcast to all robots on the team as an item to be auctioned. Robots then bid for the task based on how well-suited they are to complete it. The auctioneer then decides the winner of the auction and allocates the task to that robot. Compared to its centralized, optimization-based counterparts, auction-based methods allow each robot to bid according to their own interest. This results in a potentially sub-optimal allocation for the team, but, in return, achieves a scalable and decentralized algorithm. However, even though these algorithms are decentralized, it can require excessive communication between the agents to facilitate the auctions and thus be difficult to implement in practice~\cite{khamis2015multi}.

We seek to build a distributed and scalable way to assign roles on time-extended tasks while addressing the communication overhead of market-based task allocation algorithms. We emphasize that we seek to assign roles to the robots dynamically. This is crucial to achieving high performance as the robots need to adapt their roles online to respond to the changes in the environment \cite{gerkey2003role}. For example, in \cite{losey2020learning}, authors consider a case of speaker-listener coordination, and show assigning roles dynamically leads to far superior results than assigning static roles to robots that last for the entire duration of the task.

To do this, we use a fully game-theoretic formulation of the game and let the robots autonomously decide on their roles based on their own self-interests. Instead of explicitly coordinating the robots by assigning roles to them, we demonstrate that collaboration emerges from the strategic interactions between the robots, given that they have largely overlapping preferences. Since each robot makes decisions autonomously, this formulation does away with the communication overhead required to run the auctions. Specifically, at each time step, we let each robot construct a dynamic game that models the current state of the team, and solve the dynamic game to obtain its role for that time step. 

To allow the robots to solve the dynamics games efficiently, we apply empirical game-theoretic analysis (EGTA). Unlike dynamic game solvers \cite{fridovich2020efficient, cleac2019algames}, EGTA approximates a dynamic games with a single-shot meta-game that is easier to analyze. EGTA was first proposed in the computational economics community \cite{wellman2006methods} to analyze games that are too complex to be expressed in a closed-form. It has since seen many successes for applications in computational economics and multi-agent reinforcement learning \cite{wellman2006empirical, wah2013latency, tuyls2018generalised}. However, EGTA is usually used under scenarios where a user can afford non-real-time, offline analysis. In this work, we demonstrate through experiments that it can also be used by robots to plan for roles in real-time applications.

\emph{Statement of Contribution:}
Our contribution is three-fold. First, we present a way to formulate the role assignment problem as a dynamic game. Second, we introduce a way to apply EGTA to planning role assignments for robots. Lastly, we empirically verify our results on a simulated environment and show that our method produces role assignments that gives rise to collaborative behavior.

\emph{Organization:}
The remainder of the paper is organized as follows.
Section \ref{sec:Preliminaries} provides a concise overview of relevant concepts from game theory and section \ref{sec:Problem Formulation} introduces our game-theoretic formulation of the role assignment problem.  
Section \ref{sec:Method} describes a role-controller based on game-theoretic planning.
In Section \ref{sec: Case Study}, we illustrate the feasibility of our approach by applying it to a simulated environment on a collaborative transport task.
We conclude in Section \ref{sec: Conclusion} and suggest several directions for future work.
\section{Preliminaries}
\label{sec:Preliminaries}

\subsection{Normal-Form Games}
Normal-form games model single rounds of interactions between agents. In normal-form games, agents choose their respective strategies simultaneously. The outcome of this interaction depends on every agent's chosen strategy. Agents then receives a cost that reflects how they like this particular outcome. For example, rock-paper-scissors, where two players choose their strategies simultaneously, can be formulated as a normal-form game.

Formally, a normal-form game $G$ with $N$ players includes
\begin{itemize}
    \item a set of players $P=\{1,...,N\}$,
    \item a family of strategy sets $S = \{S^i\}_{i=1}^{N}$, and
    \item a set of payoff functions $c^i: S^i \times \Vec{S}^{-i} \rightarrow \mathbb{R}.$
\end{itemize}
The set $S^i$ denotes the set of strategies player $i$ can choose from. For example, in the case of rock-paper-scissors, $S^1=S^2=\{rock, paper, scissors\}$. A tuple built from concatenating the strategies chosen by each player $\vec{s} = (s^1, ..., s^N)$ is called a strategy profile and fully specifies the outcome of a game. We denote by $s^{-i}$ the profile found by excluding player $i$ from the profile $\vec{s}$, i.e. $s^{-i} = (s^1, ..., s^{i-1}, s^{i+1}, ..., s^N)$. We denote by $\vec{S}^{-i}$ the set of all such $s^{-i}$. The cost for player $i$ under profile $\vec{s}$ is found with the cost function at $c^i(s^i, \vec{s}^{-i})$.

In addition to playing deterministically, a player $i$ can also play stochastically by specifying a distribution over her strategy set $S^i$. This distribution $\sigma^i$ is called a mixed strategy for player $i$. For example, a good mixed strategy for rock-paper-scissors is to play a uniform distribution over the three strategies. Let $\vec{\sigma} = (\sigma^1, ...,\sigma^N)$ be a mixed-strategy profile, and $\vec{\sigma}^{-i}$ be the mixed strategies of the other $n-1$ players similar to $s^{-i}$ above. We define the expected cost to player $i$ playing a deterministic strategy $s^i$, against all other players playing according to the mixture $\vec{\sigma}^{-i}$, as
\begin{align}
c^i(s^i, \vec{\sigma}^{-i}) = \mathbb{E}_{\vec{s}^{-i} \sim \vec{\sigma}^{-i}} c^i(s^i, \vec{s}^{-i}), \label{eq:expected-cost-1}   
\end{align}
and her cost playing the mixed strategy $\sigma^i$ against $\vec{\sigma}^{-i}$ as
\begin{align}
c^i(\sigma^i, \vec{\sigma}^{-i}) = \mathbb{E}_{s^i \sim \sigma^i} c^i(s^i, \vec{\sigma}^{-i}). \label{eq:expected-cost-1}   
\end{align}

\subsection{Solving Normal-form Games}
\label{sec:prelim-learning-dynamics}
One of the most widely-used solution concepts for normal-form games is the Nash equilibrium. A strategy profile $\vec{\sigma}$ is a Nash equilibrium if
\begin{align}
    c^i(\sigma^i, \vec{\sigma}^{-i}) \leq c^i(s^i, \vec{\sigma}^{-i}),\; \forall s^i \in S^i
\end{align}
Nash equilibrium can be viewed as a prediction of the outcome of the game or as a good strategy to play against other rational players. In this work, we take the latter interpretation and solve for the Nash equilibria of a game to produce role assignments. Since solving for exact Nash equilibria is difficult \cite{chen2006settling}, one can instead approximate the Nash equilibrium of a game using learning dynamics like replicator dynamics \cite{cressman2014replicator} or fictitious play \cite{brown1951iterative}. These learning dynamics are iterative algorithms that converge to Nash equilibria under certain technical conditions on the game. Although the conditions might not always apply for the games that we are interested in analyzing in this paper, empirically, they find strategy profiles that have low regrets.

\subsection{Dynamic Games}
\label{sec:prelim-dynamic-games}
Dynamic games extend normal-form games to model strategic interactions over a dynamical system. An $N$-player discrete-time dynamic game with horizon $T$ is given by a dynamics $f$ and a set of stage-additive costs $\{J^i\}_{i=1}^{N}$. We use $x_{t} \in \mathbb{R}^n$ to denote the joint state of the system at step $t$. Note that $x_t$ captures the state of all agents at that time step and is thus not superscripted. $u^i_{t}: \mathbb{R}^n \rightarrow \mathbb{R}^{m^i}$ is the state feedback controller used by agent $i$ at time $t$. To simplify the notation, we write $u^{1:N}_t = (u^1_t, ..., u^N_t)$ as the vector of all agents' controllers at time $t$. $f$ and $J^i$ can now be written as
\begin{align}
    x_{t+1}=f_t(x_{t}, u^{1:N}_{t}),\; t=1,...,T, \label{eq:game-dynamics}
\end{align}
and
\begin{align}
    J^i(u^{1:N}_t) = \sum_{t=0}^{T}{g^i_t(x_t, u^{1:N}_t)},\;t=1,...,T.
    \label{eq:cost-functional}
\end{align}
where $g^i_t$, is a stage cost function that captures the preferences of agent $i$ over the state and controls at step $t$.

Dynamic games are general formulations that can model highly complex strategic interactions on dynamical systems. However, they are in general difficult to solve. Recent methods \cite{cleac2019algames, fridovich2020efficient} have proposed ways to approximate solutions to dynamic games. However, they are only applicable to games that have continuous dynamics and costs, and thus cannot be used to analyze games with complex dynamics. Instead, we leverage a robot's knowledge about the set of possible roles to apply empirical game-theoretic analysis analyze these dynamic games.

\subsection{Empirical Game Theoretic Analysis}
\label{sec:prelim-EGTA}
Empirical game-theoretic analysis (EGTA) \cite{wellman2006methods, tuyls2018generalised} was proposed in the computational economics community as a data-driven approach to analyze highly complex games that are difficult to solve otherwise. It does so by approximating the complex game with a small normal-form game, called the \textit{meta-game}, that only includes strategies of interest to the analysis. In the case of dynamic games, the meta-game reduces the complexity of the analysis by only considering a small set of controllers relevant to the analysis, instead of searching for a solution over all possible controllers as a dynamic game solver does. Next, we explain how to construct a meta-game from a dynamic game given the set of controllers of interest.

Consider an $N$-player dynamic game with dynamics $f$ and costs $J=\{J^i\}_{i=1}^{N}$. Assume that there is a given a set of controllers of interest $\Gamma=\{\Gamma^i\}_{i=1}^{N}$, where each $\gamma^i \in \Gamma_i$ is a sequence of state-feedback controllers for player $i$, $\gamma^i=(\gamma^i_1, ..., \gamma^i_T)$. Together, ($f$, $J$, and $\Gamma)$ induce a unique $N$-player meta-game $\hat{G}=(\hat{P}, \hat{S}, \hat{c}^{1:N})$, where
\begin{itemize}
    \item $\hat{P}=\{1,...,N\}$ is the same set of players,
    \item $\hat{S}$ is the meta-strategy set, and
    \item $\hat{c^{1:N}}$ are the meta-payoff functions.
\end{itemize}
Here, each meta-strategy corresponds to a unique controller of interest, i.e. there exist a bijection $\mathcal{M}: \hat{S}^i \rightarrow \Gamma^i$. The meta-cost for player $i$, $\hat{c}^i(\hat{s}^i, \vec{\hat{s}}^{-i})$, under the profile $\vec{\hat{s}}$ in the meta-game then corresponds to her cost in the dynamic game if all players use their corresponding controllers $\mathcal{M}(\hat{s}^j),\; j=1,...,N$. With a slight abuse of notation, we write
\begin{align}
    \hat{c}^i(\hat{s}^i, \vec{\hat{s}}^{-i}) =
    J^i(\mathcal{M}(\hat{s}^{1:N})).
\end{align}
One can then approximate the equilibria of a meta-game using learning dynamics (See Section \ref{sec:prelim-learning-dynamics}).

Several remarks are in place. First, one might wonder where such a set of controllers $\Gamma$ comes from. EGTA methods usually tackle this by having an double-oracle-like  \cite{mcmahan2003planning} outer loop around the aforementioned process that discovers interesting controllers that can be added to the meta-game \cite{lanctot2017unified}. In this work, we assume that we are given $\Gamma$ as descriptions of the behavior of each role on our robotic team.

Secondly, we note that to construct the meta-game, one has to obtain the meta-payoff $\hat{c}^i$. However, the payoff does not have to be found analytically. If one has a simulator that captures $f$ and $J$, then one can obtain $\hat{c}$ by simulating the system. Thus, this method can still be applied when the system dynamics is not differentiable or cannot be written down in closed-forms.

\subsection{Solving Empirical Games}
After the payoffs have been obtained, we apply replicator dynamics (RD), which is a system of ordinary differential or difference equations whose fixed points are ESS's. In a role-symmetric game, the dynamics is given for each role $r$ as
\[ \dot{\vec{\sigma}}^r_i = \vec{\sigma}^r_i(u(s^r_i, \vec{\sigma}) - \phi) \quad i=1,...,|S^r| \]
where $\phi = \sum_{i=1}^{|S^r|}{\vec{\sigma}_i u(s^r_i, \vec{\sigma})}$ denotes a weighted average payoff. At every time step, RD increases the probability of strategies that lead to payoffs higher than average and decreases the probabilities of those whose payoffs are lower than average. To find an ESS, our algorithm randomly initializes a mixed-strategy and iteratively applies the RD update until convergence.
\section{Role Assignment as a Meta-Game}
\label{sec:Problem Formulation}
We now explain how to formulate role assignment as a dynamic game. Consider the scenario with a team of robots working together to achieve a goal. The robots need to coordinate themselves throughout the duration of the task. We model the system of robots as a dynamical system. The dynamics of the system $f$ is defined the same way as in Equation \ref{eq:game-dynamics}, where robots share a joint state and can all assert control influence over this shared state.

Instead of modeling the goal of the team as one common function of the joint state shared by all the agents, we allow each robot's objective to differ from the team objective. This is often the case in human teams, where all members on the team collaborate to achieve a common goal, but might each have a preference on how the goal should be achieved or how to contribute to the team. We argue that allowing the robots to have different preferences makes it easier to model heterogeneous teams, as different robots have different abilities and might have different preferences as well. We model the preferences of the robots as a set of cost functions $J$, same way as defined in Equation \ref{eq:cost-functional}. We note that this formulation can still capture the case where all robots share the same objective if one sets all $J^i$ to be equal to the team objective; thus this is a strictly more general formulation. However, in the cases where the robots' preferences are not identical, there is no longer a single objective that we can optimize to obtain the optimal controller for the team. Instead, a game-theoretic setting naturally arises. We have a dynamic game uniquely defined by $f$ and $J$.

In addition to the system dynamics $f$ and the cost functionals $J$, we assume that we are given a set of possible roles to assign to the robots. The roles are modeled as a set of controllers $\Gamma$, defined in the same way as in Section \ref{sec:prelim-EGTA}. Each controller $\gamma^i \in \Gamma^i, \gamma^i: \mathbb{R}^n \rightarrow \mathbb{R}^{m^i}$ characterizes the behavior of a certain role in this game. For example, in the case of robot soccer, the set of roles (goalie, attacker, defender, etc) would each have one corresponding controller in $\Gamma$. Note that each robot $i$ has a different set of controllers $\Gamma^i$ that it can choose from. This allows us to account for the heterogeneity of the robots, where certain roles can only be assumed by a subset of the robots but not others.

Given $f$, $J$, and $\Gamma$, we can induce a meta-game $\hat{G}$ from $f$, $J$, and $\Gamma$ (the process is described in Section \ref{sec:prelim-EGTA}). By solving for the Nash equilibrium of $\hat{G}$, we find a strategy profile $\vec{\hat{\sigma}}$ that specifies, for each robot, a probabilistic distribution of the roles it should take to minimize its own cost. If the robots' preferences have a reasonable amount of overlap, as should be the case in a collaborative setting, we argue that this is a reasonable assignment that balances the robots' own interests and that of the team. We demonstrate this empirically in our case study.

To be able to build an algorithm that is fully decentralized, we further assume that each robot knows the system dynamics and the preferences of the other agents. This is a rather strong assumption, but both the system dynamics and the preferences of the other agents can be learned through system identification \cite{ljung1999system} and inverse optimal control \cite{levine2012continuous, ziebart2008maximum}, respectively. With this assumption, each robot can construct its own copy of the meta-game $\hat{G}$ and solve it to get its equilibrium strategy. By doing this, each robot can plan for its sequence of roles in fully distributed fashion.
\section{Decentralized Dynamic Role Assignment}
\label{sec:Method}
The formulation of role assignment as a meta-game in the last section already gives us a way to assign static roles to robots, i.e. the role assigned to a robot lasts for the entire duration of the task. However, it is usually beneficial for the robots to have the ability to switch roles online and dynamically determine the best role to take at a particular time step. For example, in \cite{losey2020learning}, the authors consider a case where two robots are moving through an environment but each only observes half of all the obstacles. They considered two roles, speaker and listener, and showed that the optimal performance is when the robots can switch roles infinitely often and take turns to communicate the location of the obstacles in their field of view. This intuition generalizes to many other tasks where agents have to adapt their roles to a changing environment. In this section, we show how to achieve this by extending the formulation in the last section.

Now, instead of one role, we seek to solve for a sequence of roles that specifies how a robot should adapt its behavior online. To do this, we adapt the way the meta-games are constructed. Consider the case where we allow the robots to switch roles $k$ times for the duration of the task. Between each role switch, a robot maintains a same role for $T_{rs}$ steps. To solve for the optimal sequence of $k$ roles to take, we construct a meta-game $\hat{G}$, where the meta-strategy set $\hat{S}^i=(\Gamma^i)^{k}$ for all robots $i$. That is to say, each meta-strategy $\hat{s}^i$ for a robot $i$ corresponds to a sequence of $k$ controllers $(\gamma^i_t, \gamma^i_{t+T_{rs}}, ..., \gamma^i_{t+kT_{rs}})$ that fully specifies how the roles of robot $i$ changes in the next $k\,T_{rs}$ steps. The meta-cost function is then obtained by simulating these sequences of controllers forward. In practice, we set $T_{rs}$ and $k$ as parameters and run this algorithm in a receding-horizon fashion, where we construct a meta-game $\hat{G}_t$ at every role switch ($t=m\,T_{rs}, m=0,1,...$). A robot would then take the role found in the equilibrium strategy until the next role switch at $t+T_{rs}$. In this case, $T_{rs}$ controls the frequency of role switches and $k$ controls the depth of simulation when planning for each role switch. As mentioned before, since we assume that all robots know $f$, $J$, and $\Gamma$, each robot can construct its own copy of $\hat{G}_t^i = \hat{G}_t$. Thus, our dynamic role controllers are also fully distributed.
\section{Case Study: Collaborative Transport}
\label{sec: Case Study}
We experimented our method on a 2D simulation environment (Figure \ref{fig:openloop}), where two robots try to collaboratively transport an object while avoiding collision with human agents. Below, we first give an overview of the our environment and roles considered. Then, we present the result trajectory generated by our role assignment controller.

\subsection{Collaborative Transport Environment}
\subsubsection{Dynamics}
The two robotic agents start in the bottom-left corner and are attached by an object that they want to carry to the top-right corner of the environment. The object-robot system is modeled as a one-dimensional rod, where the robots assert forces on the two ends to control the positional and angular acceleration of the system. We denote the positions of the two agents as $\vec{x}^1N = (x^{1}, y^{1})$ and $\vec{x}^2 = (x^{2}, y^{2})$. We model the state of the system as the midpoint of the rod $\vec{x}_r = \frac{1}{2}\vec{x}^1 + \frac{1}{2}\vec{x}^2$ and its angle w.r.t. the horizontal axis $\alpha$. At each step, robot $i$ asserts a force $u^i=(u^{i}_{x}, u^{i}_{y})$ on its end of the rod. The system dynamics is a double integrator given as
\begin{align}
\ddot{\vec{x}}_r &= \sum_{i} u^{i}\\
\ddot{\alpha}_i &= \sum_{i} u^{i} \times (\vec{x}^i-\vec{x}_b)
\end{align}
Two human agents in the environment try to navigate towards their respective destinations. One moves horizontally from left to right, and the other in the diagonal direction from bottom-right to top-left. The human agents in the environment are modeled as double integrators. They are driven by a potential field, where they are repulsed by the robotic agents and attracted by their respective destinations. Denote a human agent's position as $\vec{x}^h$ and its destination as $\vec{x}^{gh}$, at any point in time, it's dynamics is given by 
\begin{align*}
 \ddot{\vec{x}}^h = p^h_1 \frac{\vec{x}^{gh} - \vec{x}^h}{ ||\vec{x}^{gh} - \vec{x}^{h}||^2 } + p^h_2 \sum_{i=1}^2 \frac{\vec{x}^h - \vec{x}^{i}}{ ||\vec{x}^h - \vec{x}^{i}||^2 },
\end{align*}
where $p^h_1$ and $p^h_2$ controls the strength of the potential field. The values of $p^h_1$ and $p^h_2$ are specified in a way such that naive role controllers can easily lead the robot team to collide with the human agents. We note that humans are part of the environment and not controllable, but the robots are aware of the way the humans move and can therefore predict how the human would react to their actions.

\subsubsection{Cost Structure}
The goal for the robots are to arrive at their respective destinations while avoiding collision with the human agents and saving control effort. Formally, the stage-cost of robot $i$ at time $t$ is modeled as a sum of three components.
\begin{align*}
  g^i_t (x_t, u^{(1:2)}_t) = p^i_1 + p^i_2 \frac{1}{d_{h,t}} + p^i_3 ||u^i_t(x_t)||
\end{align*}
where $d_{h,t}$ is the minimal distance from any point on the rod to any human agent. $p^i=(p^i_1, p^i_2, p^i_3)$ controls the weighting between the three different types of costs.

We note here that each robot only cares about the size of its own control input but not that of its counterpart. This means that a robot wants to do as little work itself as possible, but still achieve the goal of transporting the object. Thus, there is an incentive for it to be a free rider and leave the heavy-lifting to the other robot. Exactly how much a robot cares about the team objective versus its own control input can be controlled by tuning the values of $p^i$.

\subsubsection{Roles Considered}
Borrowing intuition from collaborative transport in real life, we consider two roles: leader and follower. The goal of the leader is to guide the object to the destination by pulling it towards its destination. The controller for the leader is a sum of three components.
\begin{align}
    u_{leader} = (1-w) u^{g} + w u^{o} + u^{ca}
\end{align}
where $u^g, u^o, u^ca$ are respectively the positional, orientation, and collision avoidance controls. The positional control is in the direction of the goal position. The orientation control tries to adjust the angle of the rod to be horizontal. The collision avoidance control is found by a potential field approach and pushes the robot away from the human agents. $w$ is a parameter that controls the weighting between orientation and positional control and goes to one as the agent gets closer to the goal position.

The follower tries to aid the leader by mimicking the leader's control. We model the control of the follower as
\begin{align}
    u_{follower} = w \beta u_{leader} + (1-w) u^{o} + u^{ca}
\end{align}
where $u^o$ and $u^{ca}$ are found the same way as the leader. $ 0 < \beta < 1$ is a multiplier that specifies to what extent is the follower copying the leader's control. This is used to model the fact that usually, when two people move an object together, the leader expands more energy than the follower.

A robot can choose to be a leader or a follower at any point in the duration of the task, and is allowed to switch their roles online. Note again that the two robots have two largely overlapping but slightly different objective, in that they both want to arrive at the destination and avoid colliding with the humans, but they are both selfish about their own control efforts.

\subsection{Results}
\begin{figure*}[t]
    \centering
    \includegraphics[width=0.9\linewidth]{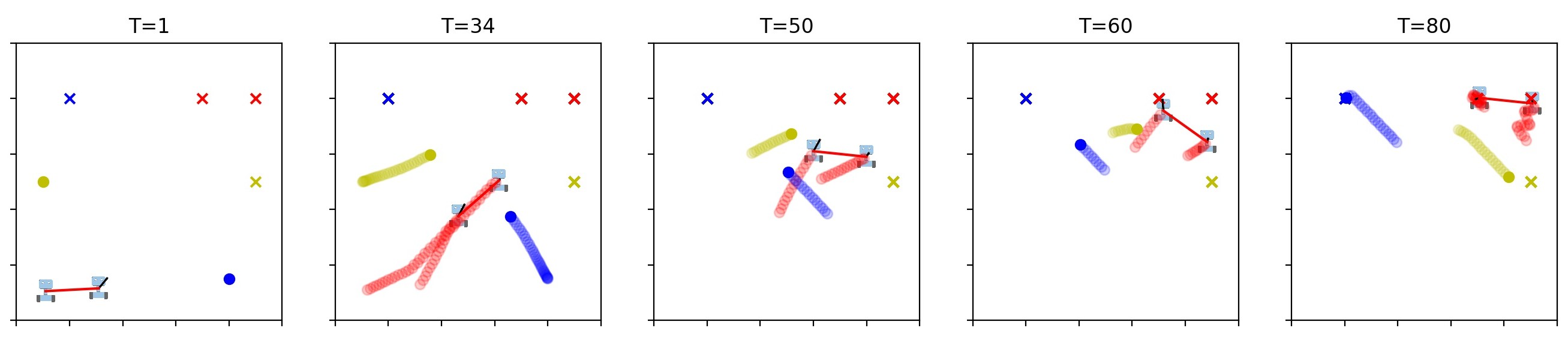}
    \caption{\textbf{Legend}: The robots (in light blue, with red trails) try to move an object (denoted by the red rod connecting two robots) from the bottom-left corner to the top-right corner. Their destinations are denoted by the red crosses. The two human agents (denoted by yellow and blue dots) try to navigate towards their respective destinations (denoted a cross of that color). \textbf{Open-loop Trajectory}: The robots reached their destination within the time limit and managed to avoid collision with the human agents. The open-loop plan is divided into three periods. During the first period (step 1-33), the right robot leads, resulting in a upward tilt of the rod. Their roles switch at step 34, when the left robot starts to lead to avoid collision with the human agents.}
    \label{fig:openloop}
\end{figure*}
\subsubsection{Open-loop}
We set the time horizon to $T=100$ and ran our controller under two specifications. First, we ran the controller in an "open-loop" fashion, where we solved for a role sequence without online replanning. In this case, we set the number of role-switches to $k=3$ and evenly divided the time horizon into equal length, i.e. a robot has the option switch its role at steps $1$, $34$, and $67$. It keeps the same role in between any role switches. To solve for its sequence of roles, the robot constructs a corresponding meta-game at the beginning of each role-switch with $k=3,T_{rs}=33,$ and $\Gamma^1=\Gamma^2=\{u_{leader}, u_{follower}\}$ (The algorithm is described in Section \ref{sec:Method}.

The equilibrium we found for this meta-game is a pure (i.e. deterministic) strategy profile, where the robot starting on the left follows in the first and third period and leads in the second. The robot starting on the right follows in the second period and leads in the first and third. We plot the resulting trajectory in Figure \ref{fig:openloop}.

First, we note that the robots succeed in their task. They arrive at their destination within the time limit and avoided collision with both human agents. Secondly, we see there is exactly one leader and one follower in each of the three time periods, which shows that our algorithm gives rise to coordination among the two agents, even though we did not enforce any constraints on collaboration. Lastly, we reiterate that by allowing each robot to only plan for itself, our algorithm generated this sequence of role assignments in a fully decentralized fashion.

\subsubsection{Closed-loop}
\begin{figure*}[t]
    \centering
    \includegraphics[width=0.9\linewidth]{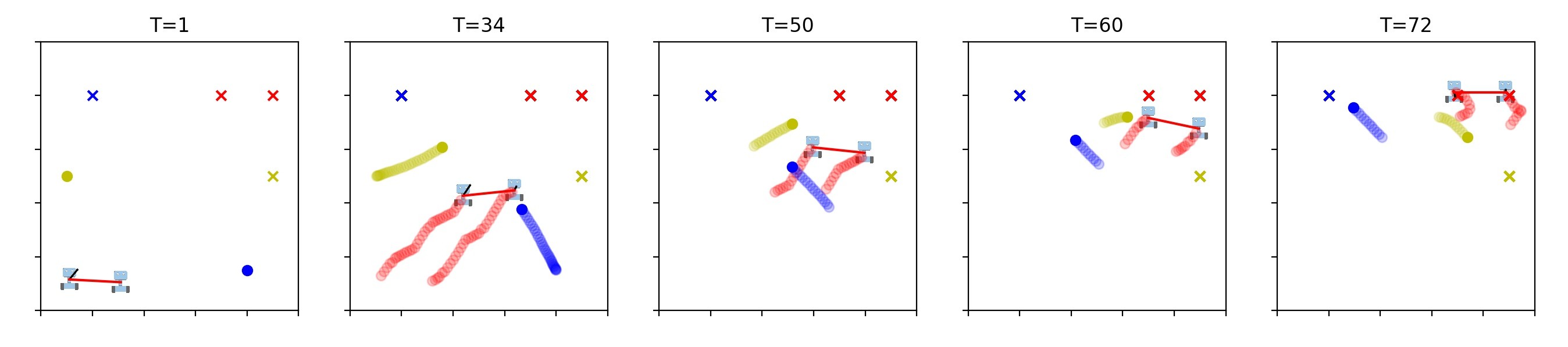}
    \caption{\textbf{Legend}: See caption for Figure \ref{fig:openloop}. \textbf{Closed-loop plan}: The robots again reach their destinations without collision with the human agents. Note that, by frequently switching roles, they were able to keep the rod level during the most parts of the episode, thus minimizing their control efforts}
    \label{fig:closedloop}
\end{figure*}
We now move on to present the results of our MPC controller. In this case, we allow the agents to switch roles every $T_{rs}=5$ time steps. For each replan, the robots looks ahead for the next $k=3$ role switches, i.e. $15$ time steps. Note that this is a significantly shorter planning horizon than that of the open-loop controller. This allows us to construct the meta-games faster, thus reducing the planning time for each role-switch. We show the trajectory of the closed loop plan in Figure \ref{fig:closedloop}.

\begin{table}[ht]
    \centering
    \begin{tabular}{c|c|c|c}
        \hline
        & Random & Open-loop & Closed-loop\\
        \hline
        player 1 cost & 1093.8  & 747.4 & 676.46\\
        player 2 cost & 1053.6 & 812.17 & 610.59\\
        Team cost & 729.85 & 442.36 & 462.08\\
        \hline
    \end{tabular}
    \caption{Cost Comparison for Different Methods}
    \label{tab:cost}
\end{table}

We note that the robots controlled by the closed-loop controller also successfully complete the task and avoids collision with human agents along the way. In fact, the costs for both agents in this trajectory are lower than their costs in the open-loop plan. They attain this reduction in cost by switching roles more often, and thus achieving maneuvers otherwise not possible in the open-loop plan.

We compare the costs of the agents for both open-loop and closed-loop settings in Table \ref{tab:cost}. In the same table, we also include the average cost achieved by a random controller that assigns a role to each robot uniformly at random. Note that a regular task allocation algorithm does not apply to our problem as the two robots have different costs, and are not included in the comparison. We see that closed-loop controller performs better than open-loop controller in terms of the cost for both agents. However, if we focus on the team objective (that is, the sum of time penalty and collision avoidance penalty but not each agent's control penalty), the closed-loop controller is less efficient than the open-loop plan on the team level. This suggests that our controller might be slightly suboptimal in terms of the team objective. However, by sacrificing a small amount of efficiency, we achieved the fully distributed algorithm. We also note that both the open-loop and the closed-loop controller achieved superior performance than the random controller.

\subsubsection{Computation Time and Scalability}
The cost of using the EGTA method can be roughly broken down into two parts. The first part of the computation cost comes from constructing the meta-game. Here, one needs to simulate all profiles to obtain their corresponding payoffs to construct the empirical game. The second part of the cost comes from solving the equilibria of the empirical game, usually through learning dynamics (Section \ref{sec:prelim-learning-dynamics}.

For moderately-sized games, the first can be addressed through parallelism. This is because the profiles can be simulated separately. The simulations can thus easily be parallelized. The performance of the second part of computation cost depends on the particular learning dynamics used. In the experiments, we observe that it takes our implementation around $1.06$ seconds to plan for each role switch. For cases where robot do not need to switch roles frequently, our algorithm can be directly applied. For tasks that need more frequent role switches, one can limit the number of iterations for the learning dynamics to reduce their running time.

\begin{remark}
We note that EGTA methods are highly scalable when the agents are \textit{symmetric} \cite{cheng2004notes}. By exploiting symmetric structures \cite{wiedenbeck2012scaling, wellman2005approximate, wiedenbeck2018regression, sokota2019learning}, EGTA methods can analyze games with up to 100 agents. Symmetric structures, however, are rare in robotic problems (the intuition is that, most of the times, two robots cannot have the exact same state within the same environment). As a next step, we plan to study how we can leverage the structures in dynamic games to relax the symmetric assumption, which could potentially lead to a dramatic increase in the scalability of of algorithm.
\end{remark}
\section{Conclusion}
\label{sec: Conclusion}
In this work, we proposed a way to apply empirical game-theoretic analysis to solve the problem of role assignment that generalizes to heterogeneous teams. We introduced a way to formulate the role assignment problem as dynamic game and presented a fully decentralized algorithm for assigning roles to robots in dynamic tasks. We evaluated our algorithm in a 2D simulation environment and demonstrated that our algorithm leads to the emergence of coordination between robots in a way that balances their own preferences and the overall goal of the team.

This method of applying EGTA in a receding-horizon fashion is by no means limited to finding role assignments. In fact, we think it has the potential to be applied in any multi-agent planning problems where the user knows multiple controllers of interest a priori. One example we consider is that of equilibrium selection. Current dynamic game solvers like \cite{cleac2019algames,fridovich2020efficient} only solve for local pure-strategy equilibria, which are not guarantee to be unique. In the case where several equilibria can be found by the solver, we think that our tool can be applied to analyze which equilibrium strategy a robot should take on.

\newpage
\bibliographystyle{IEEEtran}
\bibliography{references}

\end{document}